\documentclass[aps,prd,floatfix,noshowpacs,twocolumn,10pt,showkeys]{revtex4-1}

\usepackage{amsmath}
\usepackage{amssymb}

\usepackage{amsfonts}
\usepackage{mathrsfs}

\usepackage{graphicx}
\usepackage{dcolumn}
\usepackage{bm}
\usepackage{epsfig}
          \usepackage{epstopdf}
\usepackage{enumerate}
 \usepackage{float}
\usepackage{amssymb}

\usepackage{color}

\usepackage[colorlinks=true,linkcolor=darkblue,citecolor=darkblue,urlcolor=darkblue,breaklinks=true]{hyperref}

\usepackage[hyphenbreaks]{breakurl}
\definecolor{darkblue}{rgb}{0,0,0.7}

\def\be{\begin{equation}}
\def\ee{\end{equation}}
\def\bea{\begin{eqnarray}}
\def\eea{\end{eqnarray}}
\def\bfl{\begin{flushleft}}
\def\efl{\end{flushleft}}
\def\bfr{\begin{flushright}}
\def\efr{\end{flushright}}
\def\bc{\begin{center}}
\def\ec{\end{center}}
\def\ben{\begin{enumerate}}
\def\een{\end{enumerate}}
\def\bit{\begin{itemize}}
\def\eit{\end{itemize}}

\def\dzn{,\kern-0.1em,}

\def\d#1{{#1\kern-0.4em\char"16\kern-0.1em}}
\def\D#1{{\raise0.2ex\hbox{-}\kern-0.4em 31}}
\def\e{\text{e}}
\def\i{\text{i}}
\def\d{\text{d}}
\def\Lan{\langle}
\def\Ran{\rangle}

\newcommand {\apgt} {\ {\raise-.5ex\hbox{$\buildrel>\over\sim$}}\ }
\newcommand {\aplt} {\ {\raise-.5ex\hbox{$\buildrel<\over\sim$}}\ }

\begin{document}




\title{Dynamics of Frenkel excitons in pentacene}


\author{Sonja Gombar}
\affiliation{Department of Physics, Faculty of Sciences, University of Novi Sad, Trg Dositeja
 Obradovi\' ca 4, Novi Sad, Serbia}
\author{Petar  Mali}
\affiliation{Department of Physics, Faculty of Sciences, University of Novi Sad, Trg Dositeja
 Obradovi\' ca 4, Novi Sad, Serbia}
\email{petar.mali@df.uns.ac.rs}
\author{Milan Panti\' c}
\affiliation{Department of Physics, Faculty of Sciences, University of Novi Sad, Trg Dositeja
 Obradovi\' ca 4, Novi Sad, Serbia}
 \author{Milica Pavkov-Hrvojevi\' c}
\affiliation{Department of Physics, Faculty of Sciences, University of Novi Sad, Trg Dositeja
 Obradovi\' ca 4, Novi Sad, Serbia}
\author{Slobodan  Rado\v sevi\' c}
\affiliation{Department of Physics, Faculty of Sciences, University of Novi Sad, Trg Dositeja
 Obradovi\' ca 4, Novi Sad, Serbia}





\begin{abstract}
The dispersion relation for noninteracting excitons and the influence of 
perturbative correction is examined in the case of pentacene structure. 
The values of exchange integrals are determined by the nonlinear 
fits to the experimental dispersion data obtained by inelastic electron 
scattering in Phys. Rev. Lett. \textbf{98}, 037402 (2007). We obtain 
theoretical dispersion curves along four different directions in 
the Brillouin zone which possess the same periodicity as the experimental data. 
We also showed that  perturbative corrections are negligible 
since the exciton gap in dispersion relation is huge in 
comparison to exchange integrals.

\end{abstract}





\maketitle

\section {Introduction}
In the last few decades organic molecular solids have been matter of intense 
theoretical and experimental studies, due to their potential applications 
in novel organic
devices \cite{forest,jang}. The recent advances in experimental methods 
have provided detailed insight into their microscopic properties 
\cite{breakdown}. Among the energetically lowest excitations in such  
systems are Frenkel excitons, electron-hole pairs of small radius 
\cite{Agranovic}. The general theory of Frenkel excitons in molecular 
crystals is exposed in detail in \cite{Agranovicbook},
while some references on applications and progress
are \cite{SciRep,BokAgr,ExcPhysRepts,Panta}.

The method of inelastic electron 
scattering was used for
direct  measurement of the  exciton band  structure  within  the
reciprocal $\bm{a}^{*}\bm{b}^{*}$ plane of pentacene at room temperature ($T=300$K) in \cite{breakdown}. 
Results of measurements along four different directions
in the Brillouin zone were presented and, on that basis, 
the authors of \cite{breakdown} argued
that the model of noninteracting Frenkel excitons is 
inapplicable for description of pentacene (see also \cite{20K} for measurements at $T=20$K and \cite{pic} for similar experiments on picene).  
They also  suggested that the charge-transfer
(CT) excitons must be included in model Hamiltonian
of pentacene in order to achieve better agreement with experiment.
Following these experiments, a significant theoretical work was 
conducted in order to obtain properties of pentacene 
from the first principles, i.e. starting from
 many-body electron-hole Hamiltonians \cite{GatuzoPRB,GatuzoPRB2,GatuzoJPC,AnnRev,JPCLett}.

The present paper deals with the problem of obtaining exciton dispersion
in pentacene by relying on a correspondence
between Paulion Hamiltonian and anisotropic $XXZ$
Heisenberg ferromagnet. Unlike previous theoretical works, 
based on many-body  Hamiltonians containing electron and hole
creation and anihilation operators, we present calculations
based on effective Hamiltonian \cite{WQFT2,Wen,Brauner}. In other words, we start from
Frenkel excitons as low lying degrees of freedom and obtain
effective form of their interactions which are considered
to the one loop order.

Whereas our results confirm
that exciton dispersion in  pentacene can not be described
within Frenkel model alone in satisfactory manner,
they also suggest that the influence of other excitations
may not be as large as originally proposed.

The paper is organized as follows. The model Hamiltonian
and pentacene structure are  introduced
in Sec. II, while exciton dispersion in noninteracting model
is obtained in Sec. III. Finally, we discuss the effects
of exciton-exciton interactions within two-level model
in Sec. IV and V.

\section {Model Hamiltonian and pentacene structure} 
The basic Hamiltonian that governs the dynamics of excitons in two-level  
system (only one electronically excited molecular state is considered) 
is given by
\bea 
H&=&H_0+\Delta \sum_{\bm{n}}P^+_{\bm{n}}P_{\bm{n}}
-\frac{X}{2}\sum_{\bm{n},\bm{\lambda}}P^+_{\bm{n}}P_{\bm{n}+
\bm{\lambda}}\nonumber\\
&-&\frac{Y}{2}\sum_{\bm{n},\bm{\lambda}}P^+_{\bm{n}}P_{\bm{n}}
P^+_{\bm{n}+\bm{\lambda}}P_{\bm{n}+\bm{\lambda}}. \label{paulham} 
\eea
where $P^+_{\bm{n}}$ and $P_{\bm{n}}$ are standard Pauli 
operators on the site $\bm{n}$ and $X$ and $Y$ are parameters describing
hopping and interactions of excitons, respectively \cite{Agranovicbook,Tole}. 
Using the exact one to one correspondence between 
Pauli and spin operators in the case of $S=1/2$ \cite{Tjablikov}, we obtain 
anisotropic 
($XXZ$) Heisenberg Hamiltonian in external field 
\be 
H=-\frac{I^x}{2}\sum_{\bm{n},\bm{\lambda}}S^-_{\bm{n}}
S^+_{\bm{n}+\bm{\lambda}}-\frac{I^z}{2}\sum_{\bm{n},\bm{\lambda}}
S^z_{\bm{n}}S^z_{\bm{n}+\bm{\lambda}}-\mu \mathcal{H}
\sum_{\bm{n}}S^z_{\bm{n}}, \label{spinham}
\ee
where $\{\bm{\lambda}\}$ denotes vectors connecting neighboring sites, $z_1$ is
the number of nearest neighbours and
\bea
I^z = Y, \quad
I^x = X, \quad
\mu \mathcal{H} = \Delta - \frac{I^z z_1}{2}. \label{veza}
\eea
Equivalently, inverse relations are
\bea
 \Delta&=&\frac{I^zz_1}{2}+\mu \mathcal{H}, \quad
 Y=I^z, \quad 
 X=I^x, \quad \nonumber \\ 
 H_0&=&-\frac{I^zNz_1}{8}-\frac{N\mu \mathcal{H}}{2}. \label{veza2} 
\eea
Due to the isomorphism of spin and paulion Hilbert spaces on every lattice 
site and relations (\ref{veza})-(\ref{veza2}), the original problem of exciton 
dynamics governed by (\ref{paulham}) can be completely mapped on 
the equivalent effective spin model (\ref{spinham}). It should be noted 
that this correspondence is purely formal -- it will allow us
to investigate the exciton system with the help of
a vast number of existing theoretical tools 
developed for  spin systems \cite{Tjablikov,Kuntz,Averbah,Mano,Nolting,Sandvik,SSC,Hofman1,Hofman2}.
According to \cite{Agranovicbook},
Pauli Hamiltonian (\ref{paulham}), which is as we have shown here equivalent to anisotropic 
Heisenberg Hamiltonian (\ref{spinham}), can be used in description of pentacene. 
This fact will enable us to examine dispersion of noninteracting excitons as well as
the influence of their interactions in leading order (one-loop) approximation.
One should also note that, in  many practical cases, the set of neighboring sites connected with 
hopping integrals splits into several subsets, determined by the 
lattice structure and values of hopping parameters.

\begin{figure} 
\includegraphics[scale = 0.45]{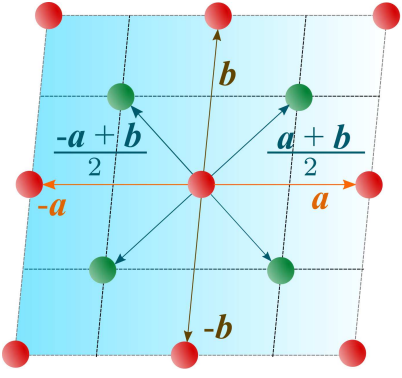}
{\caption{\label{PhaseFM} Schematic presentation of the
pentacene lattice. A pair of exchange integrals corresponds to the each set of lattice vectors  $\{\bm{a},-\bm{a}\}$,
$\{\bm{b},-\bm{b}\}$ and $\{\frac{\bm{a}+\bm{b}}{2},
\frac{-\bm{a}+\bm{b}}{2},-\frac{\bm{a}+\bm{b}}{2},-\frac{-\bm{a}+\bm{b}}{2}\}$ (see text).}}
\end{figure}

We shall analyze now the pentacene structure. A schematic sketch of the pentacene thin film lattice is shown in Figure 
\ref{PhaseFM}.
The lattice parameters within the $\bm{a}\bm{b}$ layer of the single crystal of 
pentacene are $|\bm{a}|=6.27\mathring {\mbox{A}}$, $|\bm{b}|=7.78\mathring 
{\mbox{A}}$, $ \sphericalangle(\bm{a},\bm{b})=87.8^{\circ}$  \cite{Ugao}. 
 Central motive in Figure \ref{PhaseFM} has three types of 
neighbours: two neighbours at points $\bm \lambda_1=\{\bm{a},-\bm{a}\}$ 
coupled trough exchange integral $I_1$, two neighbours at points $\bm \lambda_2 = 
\{\bm{b},-\bm{b}\}$  coupled through exchange integral $I_2$ and four neighbours 
at points $\bm \lambda_3 = \{\frac{\bm{a}+\bm{b}}{2},\frac{-\bm{a}+\bm{b}}{2},
-\frac{\bm{a}+\bm{b}}{2},-\frac{-\bm{a}+\bm{b}}{2}\}$ coupled via exchange integral $I_3$. 
As we have already noted, the transition from Pauli to Heisenberg Hamiltonian
requires anisotropic exchange integrals. Therefore, each of the mentioned exchange
integrals splits into $x$ and $z$ components: $I_j \rightarrow (I_j^x,I_j^z) $,
where $j = 1,2$ or $3$.
The results from a recent paper \cite{Mob2D}
show that the most important hopping paths in the
pentacene crystal are in the planes perpendicular to the $\bm c^*$ axis. Thus,
to a good approximation, real pentacene crystal can be effectively
described by  a two dimensional model.
Finally, by following common practice \cite{breakdown}, we shall present
numerical results for the approximate pentacene lattice, defined by the additional 
constraint $\bm a \cdot \bm b = 0$.
\bc
\ec
\section{Dispersion of noninteracting excitons}
\begin{figure} 
\includegraphics[width=\columnwidth]{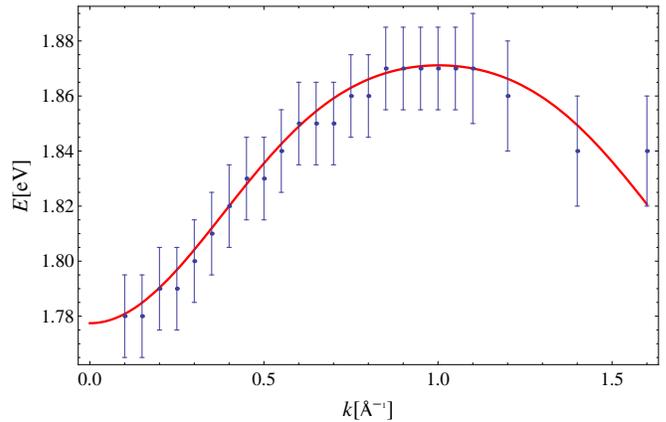} 
{\caption{\label{Fig100}
Exciton dispersion along $(100)$ direction. Experimental data are taken 
from \cite{breakdown}. Theoretical curve is obtained for: $\Delta = 1.83 \text{eV}$ \cite{183},
$I^x_1=5.7 \text{meV}$ and $I^x_3=23.4 \text{meV}$.}}
\end{figure}

\begin{figure} 
\includegraphics[width=\columnwidth]{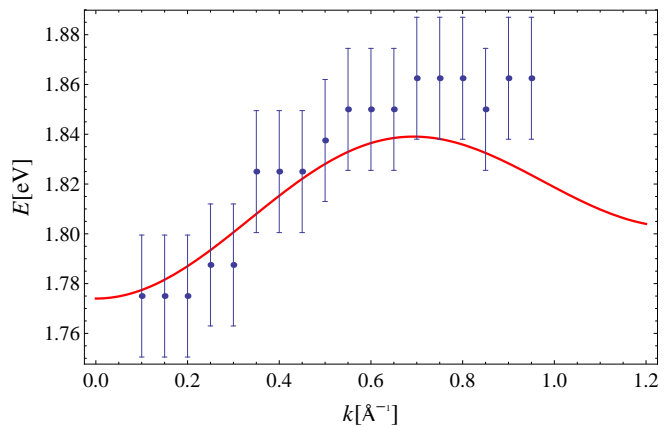}
{\caption{\label{Fig210}
Exciton dispersion along $(210)$ direction. Experimental data are taken 
from \cite{breakdown}. Theoretical curve is obtained for: $I^x_2=3.4 \text{meV}$ 
(the rest of the parameters are as in Fig. \ref{Fig100}).}}
\end{figure}

\begin{figure} 
\includegraphics[width=\columnwidth]{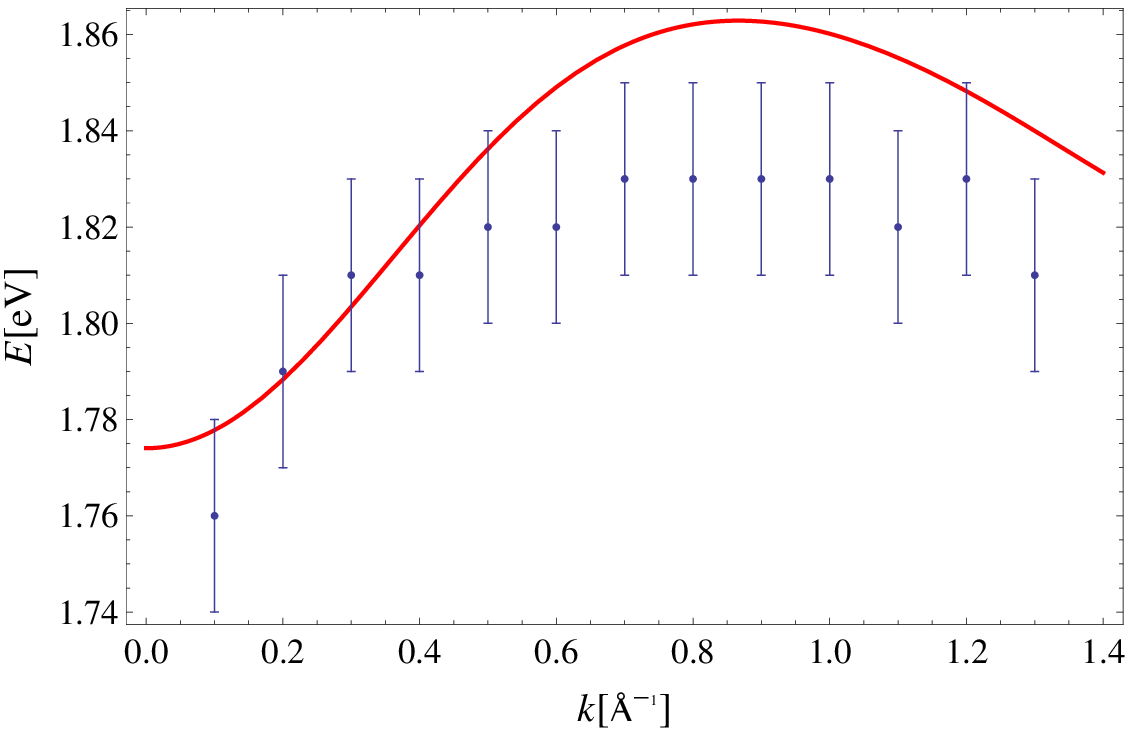}
{\caption{\label{Fig120}
Exciton dispersion along $(120)$ direction. Experimental data are taken 
from \cite{breakdown}. Parameters used for theoretical fit are 
as in Figures \ref{Fig100} and \ref{Fig210}.}}
\end{figure}

\begin{figure} 
\includegraphics[width=\columnwidth]{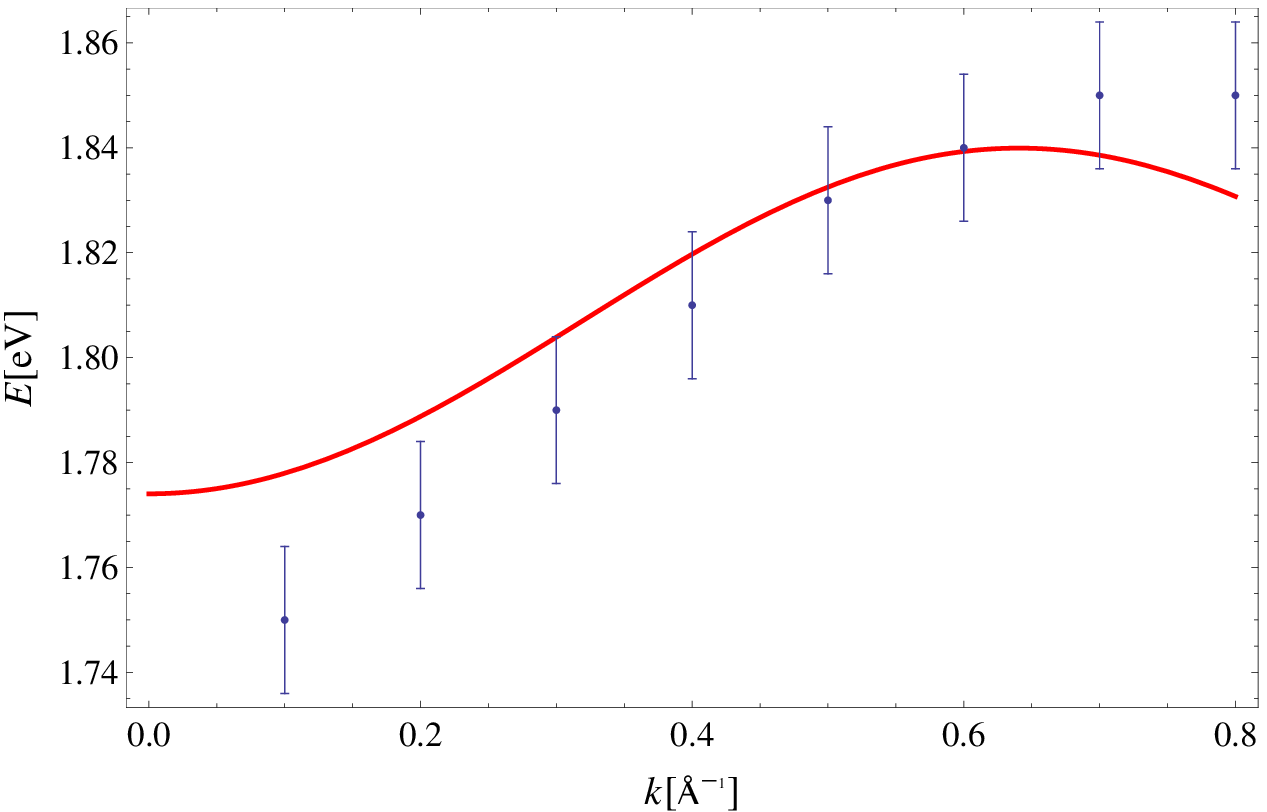}
{\caption{\label{Fig110}
Exciton dispersion along $(110)$ direction. Experimental data are taken 
from \cite{breakdown}. Parameters used for theoretical fit are 
as in Figures \ref{Fig100} and \ref{Fig210}.}}
\end{figure}

\begin{figure} 
\includegraphics[width=\columnwidth]{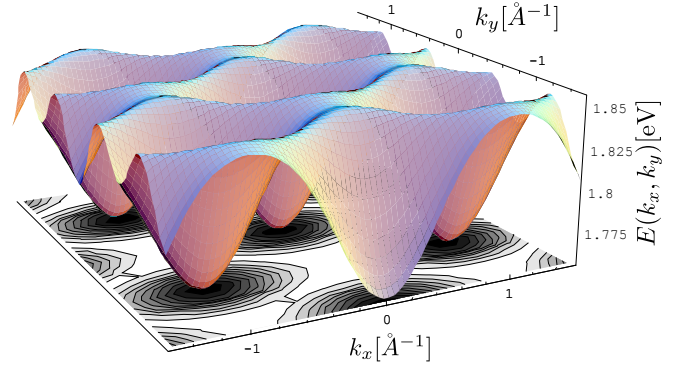}
{\caption{\label{Fig3D}
Exciton dispersion  in three dimensions. Parameters used for theoretical fit are 
as in Figures \ref{Fig100} and \ref{Fig210}. }}
\end{figure}

Bearing in mind remarks on pentacene structure from previous section,
we obtain Hamiltonian (\ref{spinham}), adapted to the 
pentacene structure,  in the Bloch approximation, 
\bea 
H&=&H'_0-\frac{1}{2}\sum_{j}I^x_{j}
\sum_{\bm{n},\bm{\lambda}_{j}}B^{\dagger}_{\bm{n}}B_{\bm{n}+\bm{\lambda}_{j}}
+\frac{1}{2}\sum_{j}I^z_{j}\sum_{\bm{n},\bm{\lambda}_{j}}B^{\dagger}_{\bm{n}}B_{\bm{n}} \nonumber\\
&+& \mu \mathcal{H}\sum_{\bm{n}}B^{\dagger}_{\bm{n}}B_{\bm{n}} \nonumber\\
&=& H'_0-\frac{I^x_1}{2}\sum_{\bm{n},\bm{\lambda}_{1}}B^{\dagger}_{\bm{n}}B_{\bm{n}
+\bm{\lambda}_{1}}+\frac{I_1^z}{2}\sum_{\bm{n},\bm{\lambda}_{1}}B^{\dagger}_{\bm{n}}B_{\bm{n}} \nonumber\\
&-& \frac{I_2^x}{2}\sum_{\bm{n},\bm{\lambda}_{2}}B^{\dagger}_{\bm{n}}B_{\bm{n}
+\bm{\lambda}_{2}}+\frac{I_2^z}{2}\sum_{\bm{n},\bm{\lambda}_{2}}B^{\dagger}_{\bm{n}}B_{\bm{n}} \nonumber\\
&-&  \frac{I_3^x}{2}\sum_{\bm{n},\bm{\lambda}_{3}}B^{\dagger}_{\bm{n}}B_{\bm{n}
+\bm{\lambda}_{3}}+\frac{I_3^z}{2}\sum_{\bm{n},\bm{\lambda}_{3}}B^{\dagger}_{\bm{n}}B_{\bm{n}}\nonumber\\
&+&\mu \mathcal{H}\sum_{\bm{n}}B^{\dagger}_{\bm{n}}B_{\bm{n}} ,
\eea
where $B_{\bm n}^\dagger \;(B_{\bm n})$ are boson creation (annihilation) operators.
The same Hamiltonian in $\bm{k}$ space has the form:
\bea \tilde{H}&=&\tilde{H}'_0-\frac{I_1^x}{2}\sum_{\bm{k}}
B^{\dagger}_{\bm{k}}B_{\bm{k}}z_1\gamma_1(\bm{k})+\frac{I_1^zz_1}{2} 
\sum_{\bm{k}}B^{\dagger}_{\bm{k}}B_{\bm{k}} \nonumber\\
&-& \frac{I_2^x}{2}\sum_{\bm{k}}B^{\dagger}_{\bm{k}}
B_{\bm{k}}z_2\gamma_2(\bm{k})+\frac{I_2^zz_2}{2} \sum_{\bm{k}}
B^{\dagger}_{\bm{k}}B_{\bm{k}} \nonumber\\
&-& \frac{I_3^x}{2}\sum_{\bm{k}}B^{\dagger}_{\bm{k}}B_{\bm{k}}
z_3\gamma_3(\bm{k})+\frac{I_3^zz_3}{2} \sum_{\bm{k}}B^{\dagger}_{\bm{k}}B_{\bm{k}}\nonumber\\
&+&\mu \mathcal{H}\sum_{\bm{k}}B^{\dagger}_{\bm{k}}B_{\bm{k}}, \eea
where $z_1 = z_2 = 2$, $z_3 = 4$ and 
corresponding geometric factors are defined by
\be \gamma_{1}(\bm{k})=\frac{1}{2}\sum_{\bm{\lambda}_{1}}\e^{\i \bm{k} \cdot \bm{\lambda}_{1}}
=\frac{1}{2}\left(\e^{\i \bm{k}\cdot \bm{a}}+\e^{-\i \bm{k}\cdot \bm{a}}\right)=\cos(\bm{k}\cdot \bm{a}) ,\ee
\be \gamma_{2}(\bm{k})=\frac{1}{2}\sum_{\bm{\lambda}_{2}}\e^{\i \bm{k} \cdot \bm{\lambda}_{2}}
=\frac{1}{2}\left(\e^{\i \bm{k}\cdot \bm{b}}+\e^{-\i \bm{k}\cdot \bm{b}}\right)=\cos(\bm{k}\cdot \bm{b}), \ee
\bea \gamma_{3}(\bm{k})&=&\frac{1}{4}\sum_{\bm{\lambda}_3}\e^{\i \bm{k}\cdot \bm{\lambda}_3}
\nonumber\\
&=&\frac{1}{4}\Big(\e^{\i \bm{k}\cdot \frac{\bm{a}+\bm{b}}{2} }+\e^{-\i \bm{k}\cdot 
\frac{\bm{a}+\bm{b}}{2} }+\e^{\i \bm{k}\cdot \frac{\bm{a}-\bm{b}}{2} }+\e^{-\i \bm{k}
\cdot \frac{\bm{a}-\bm{b}}{2} }\Big) \nonumber\\
&=& \frac{1}{2} \cos\left[\frac{\bm{k}\cdot(\bm{a}+\bm{b})}{2}\right]+\frac{1}{2} 
\cos\left[\frac{\bm{k}\cdot(\bm{a}-\bm{b})}{2}\right] \nonumber\\
&=& \cos\left(\frac{\bm{k}\cdot \bm{a}}{2}\right)\cos\left(\frac{\bm{k}\cdot \bm{b}}{2}\right). \eea
From 
\be
 \tilde{H}=\tilde{H}'_0+\sum_{\bm{k}}E(\bm{k})B^{\dagger}_{\bm{k}}B_{\bm{k}} 
\label{BlochHam}
\ee
we obtain dispersion relation
\bea E(\bm{k})&=&I_1^x\left[\frac{I_1^z}{I_1^x}-\cos(\bm{k} \cdot \bm{a})\right]+
I_2^x\left[\frac{I_2^z}{I_2^x}-\cos(\bm{k} \cdot \bm{b})\right]\nonumber\\
&+&2I^x_3\left[\frac{I^z_3}{I^x_3}-\cos\left(\frac{\bm{k} \cdot \bm{a}}{2}\right)
\cos\left(\frac{\bm{k} \cdot \bm{b}}{2}\right)\right]+\mu \mathcal{H}. \nonumber \eea
That is,
\bea E(\bm{k})&=&\Delta-I^x_1 \cos(\bm{k}\cdot \bm{a})-I_2^x \cos(\bm{k}\cdot \bm{b})\nonumber\\
&-&2I_3^x \cos\left(\frac{\bm{k} \cdot \bm{a}}{2}\right)\cos\left(\frac{\bm{k} \cdot \bm{b}}{2}\right), \label{Blochdis} \eea
where we have defined the exciton gap
\be
\Delta=I_1^z+I_2^z+2I_3^z+\mu \mathcal{H}. \label{DeltaGap}
\ee
Exciton dispersion (\ref{Blochdis}) law is plotted along $(100)$
in Fig. \ref{Fig100}. Note that the orthogonality condition $\bm a \cdot \bm b = 0$
allows us to determine 
$I^x_1 = 5.7 \text{meV}$ and $I^x_3=23.4 \text{meV}$ by fitting (\ref{Blochdis})  to experimental 
data along this direction only
(see the paper \cite{SSCDisp} for a discussion on determination of exchange
integrals from dispersion relation in a similar
context). The value $\Delta = 1.83 \text{eV}$ is
taken from \cite{183}. 
The last  parameter $I^x_2=3.4 \text{meV}$ 
is extracted from the experimental data on exciton dispersion
along $(210)$ direction (see Fig. \ref{Fig210}).
By using this set of parameters, we have plotted the exciton dispersion
along $(110)$ and $(120)$ directions and compared them to the experimental
data from \cite{breakdown}. Since we have used a single set of model
parameters, the plotted dispersion law displays the unique limit
$\Delta - I_1^x - I_2^x -2 I_3^x = 1.7741 \text{eV}$ as $|\bm k| \to 0$
for all four directions in Brillouin zone.
This is clearly seen from Figs \ref{Fig100}, \ref{Fig210}, \ref{Fig120} and \ref{Fig110}.
Finally, 3D plot of exciton dispersion
$E(k_x,k_y)$
is given in Fig. \ref{Fig3D}.

The disagreement between  dispersion of excitons predicted by the noninteracting
Frenkel model and experimental data, which is evident from  
Figs \ref{Fig100}, \ref{Fig210}, \ref{Fig120} and \ref{Fig110}, 
should be
attributed to the existence of other excitations (CT excitons) in the system according to \cite{breakdown}.
Specifically, the difference between theoretical curve and experiment was the most prominent along $(120)$
direction in \cite{breakdown}, since calculated and measured dispersion 
do not share the same periodicity.
Even though the agreement between theory and experiment is the best for
$(100)$ direction, theoretical curves presented here possess the same periodicity
as experimental  data within Brillouin zone, for all four directions.  
Thus, the influence of CT excitons may not be as large as originally
suggested in \cite{breakdown}.

The exciton dispersion in paper \cite{breakdown} is given by
\bea E(\bm{k})&=&E_0+t_{\bm{a}}\cos(\bm{k}\cdot \bm{a})+t_{\bm{b}}
\cos(\bm{k}\cdot \bm{b})\nonumber\\
&+&2t_{\bm{a}\bm{b}}\cos\left(\frac{\bm{k} \cdot \bm{a}}{2}\right)
\cos\left(\frac{\bm{k} \cdot \bm{b}}{2}\right). \label{Blochdisprl} \eea
Since the effective mass of excitons in pentacene is large \cite{PRL2},
hopping parameters in (\ref{paulham}) are positive and small
so that corresponding Heisenberg Hamiltonian (\ref{spinham})
describes a ferromagnet.
By comparing relations 
(\ref{Blochdis}) and (\ref{Blochdisprl}) we find
\bea \hspace*{-0.5cm} t_{\bm{a}}=-I_1^x<0, \quad
 t_{\bm{b}}=-I_2^x<0 , \quad
 t_{\bm{a}\bm{b}}=-I_3^x<0. \eea
Therefore,  without fitting dispersion on experimental 
data, we conclude that $t_{\bm{a}}$, $t_{\bm{b}}$, $t_{\bm{a}\bm{b}}$ 
must be negative, which is in accordance with 
\cite{breakdown}.

\bc
\ec

\bc
\ec

\bc
\ec

\bc
\ec

\bc
\ec
\section{Perturbative corrections} 

In the previous section we saw that the model of noninteracting excitons
gives dispersion law which lies within error bars almost within entire
Brillouin zone. The question, which naturally follows this observation, is: could the
agreement between theory and experiment be improved by including the
effects of exciton-exciton interactions?
By answering to this question we could provide additional support for the hypothesis proposed
for the first time in \cite{breakdown}, according to which additional excitations
(CT excitons) need to be taken into account for correct description
of pentacene.

A careful examination of 
traditional methods for studying the effect of interactions in models
based on Pauli/Heisenberg Hamiltoninans (\ref{paulham}) - (\ref{spinham})
reveals that they
possess certain flaws. To avoid them, we employ perturbation theory
developed in \cite{Sloba,Sloba2}. The main advantage of this method is that boson 
representations of spin operators are unnecessary. In other words, 
exciton-exciton interaction, which is partially hidden in the spin 
Hamiltonian and partially in the corresponding Hilbert space, is 
explicitly given through interaction pieces of Lagrangian. 
Therefore, perturbative corrections may be calculated more 
systematically. This is extremely important for $S=1/2$ spin 
Hamiltonians, since $1/S$ is not a small parameter that can control perturbative calculations. 

As it is well known, the Lagrangian that reproduces 
Landau-Lifshitz equation is \cite{PRDLeut,WatanabePRL,Wen}
\bea
\mathcal{L}_{\text{eff}} = \Sigma \frac{\partial_t U^1 U^2 - \partial_t U^2 U^1}{1+U^3}
-\frac{F^2}{2} \partial_{\alpha} U^i \partial_{\alpha} U^i+\Sigma \mu \mathcal{H} U^3\nonumber\\ \label{EffLagr}
\eea
where two excitation fields are collected into the unit 
vector $\bm{U}:=[U^1\,\, U^2 \, \,U^3]^{\text T}
\equiv [\bm \pi (x),U^3(x)]^{\text T}$, $\Sigma = N S/V$  
and $F$ is a constant to be determined later. 

Corresponding free part of Lagrangian is
\bea
\mathcal{L}_{\text{free}} = \frac{\Sigma}{2}  
\left[ \partial_t \pi^1 \pi^2 - \partial_t \pi^2 \pi^1 \right]
+\frac{F^2}{2} \bm \pi \cdot \partial_{\alpha}\partial_{\alpha} 
\bm \pi+\Sigma \mu \mathcal{H}\bm{\pi}^2 \nonumber\\ \label{Lfree}
\eea
and interaction part up to quartic approximation 
(which is sufficient for one loop calculations) is
\be \mathcal{L}_{\text{int}}=\frac{F^2}{8}\bm{\pi}^2
\partial_{\alpha}\partial_{\alpha} \bm{\pi}^2-
\frac{F^2}{8}\bm{\pi}^2\bm{\pi}\cdot \partial_{\alpha}\partial_{\alpha} \bm{\pi}. \ee

\begin{figure} 
\includegraphics[width=\columnwidth]{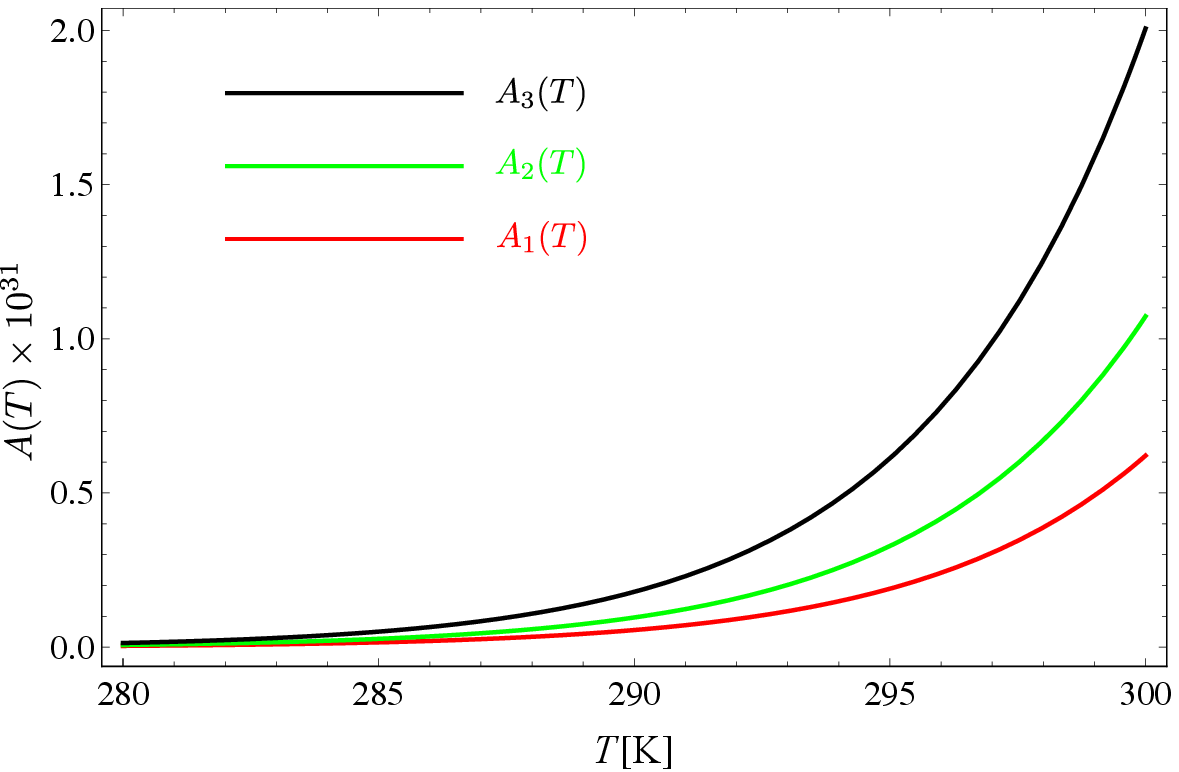}
\caption{\label{ATemp}}
Renormalizing factors $A_j(T)$ defined in (\ref{AjT}). 
\end{figure}

To obtain the Hamiltonian suitable for perturbative 
calculation, we apply canonical quantization and incorporate 
the structure of the lattice presented at Fig. \ref{PhaseFM}. 
Basically this means that we wish to construct the free 
Hamiltonian with lattice exciton fields that reproduces 
exciton dispersion (\ref{Blochdis})
\be 
H_0=-\frac{\upsilon_0}{2m}\sum_{\bm{x}}
\psi^{\dagger}\text{D}^2 \psi+\mu \mathcal{H} \upsilon_0 \label{FreeHam}
\sum_{\bm{x}}\psi^{\dagger}\psi,
\ee
where $\upsilon_0=ab$, $\psi$ and $\psi^{\dagger}$ 
satisfy canonical commutation relations for Schroedinger fields
and $\text{D}^2$ and $m$ are the discrete Laplacian and parameter defined in such a way for
(\ref{FreeHam}) to reproduce dispersion (\ref{Blochdis}).
It can be readily checked that  $\text{D}^2$ is given by
\bea
\text{D}^2  & = & \nabla^2_{(3)}
 + \frac{1}{2} \frac{|\bm \lambda_1|^2}{|\bm \lambda_3|^2} \frac{I_1^x}{I^x_3} \nabla^2_{(1)}
+ \frac{1}{2} \frac{|\bm \lambda_2|^2}{|\bm \lambda_3|^2} \frac{I_2^x}{I^x_3} \nabla^2_{(2)},
\eea
where
\bea
\nabla^2_{(j)} \phi(\bm x) &:= &\frac{4}{z_j |\bm \lambda_j|^2} \sum_{\bm \lambda_j}
 \Big( \phi(\bm x + \bm \lambda_j) - \phi(\bm x)  \Big), \nonumber \\ 
j & = & 1,2,3  \label{DiscLapDef}
\eea
are the discrete Laplacians for three sets of neighbors (see the Fig. \ref{PhaseFM})
and
\bea
m = \frac{1}{I^x_3 |\bm \lambda_3|^2} \equiv \frac{\Sigma}{2 F^2}. \label{MassDef}
\eea
Further, it is useful to introduce eigenvalues of discrete Laplacians.
They are given by
\bea
\nabla^2_{(j)} \e^{\i \bm k \cdot \bm x} = - \widehat{\bm k}^2_{(j)} \e^{\i \bm k \cdot \bm x},
\nonumber \\
\widehat{\bm k}^2_{(j)} := \frac{2D}{|\bm \lambda_{j}|^2} \Big(  1-\gamma_{j}(\bm k)  \Big)
\eea
and the exciton dispersion, obtained from (\ref{FreeHam}) by a Fourier transform, may be written as
\begin{widetext}
\bea
E(\bm k)  = \mu \mathcal{H} + \delta_1 + \delta_2 + \delta_3 
 +  \frac{I_3^x|\bm \lambda_3|^2}{2} 
\left[ \widehat{\bm k}^2_{(3)} + 
 \frac{1}{2} \frac{|\bm \lambda_1|^2}{|\bm \lambda_3|^2} \frac{I_1^x}{I^x_3}  \widehat{\bm k}^2_{(1)} 
+\frac{1}{2} \frac{|\bm \lambda_2|^2}{|\bm \lambda_3|^2} \frac{I_2^x}{I^x_3}    \widehat{\bm k}^2_{(2)} \right]
\equiv \Delta + \frac{\widehat{\bm k}^2}{2 m} \label{KSqDef}
\eea
\end{widetext}
where $\delta_j = I^z_j - I^x_j$. 
Thus, Hamiltonian (\ref{FreeHam}) is equivalent to the Bloch
Hamiltonian (\ref{BlochHam}). 
Note that the discrete Laplacian $\text{D}^2$, which defines local
changes of  the excitation fields, depends on ratios 
 $I^x_1/I^x_3$ and $I^x_2/I^z_3$.
Thus, the full symmetry of the pentacene lattice, which
reflects itself through the energies of free excitons, can be implemented
within effective model  only
by the right choice of exchange integrals.
The exciton-exciton interactions,
which modify exciton dispersion to the one loop can now be written
as (see \cite{Sloba, Sloba2})
\be
H_{\text{int}} = H_{4}^{(a)} + H_{4}^{(b)},
\ee
with
\bea
H_4^{(a)} & = &\frac{F^2}{8}  v_0 \sum_{\bm x} \bm \pi^2(\bm x)  \bm \pi(\bm x) \cdot \text{D}^2 \bm \pi(\bm x)
\label{HamEffLatt1}, \\
H_4^{(b)} & = & -\frac{F^2}{8}  v_0 \sum_{\bm x} \bm \pi^2(\bm x) \text{D}^2 \bm \pi^2(\bm x), \nonumber 
\eea
and
\bea
\psi & = & \sqrt{\frac{2}{\Sigma}} \Big[ \pi^2 + \i \pi^2  \Big].
\eea
Now we can calculate the one-loop correction to the exciton dispersion.
It is determined by the one-loop self energy which, in turn, may
be calculated by the diagrammatic rules introduced in \cite{Sloba,Sloba2}.
In short, we see from (\ref{HamEffLatt1}) that the excitons are coupled
derivatively so that internal and external lines on Feynman diagrams
carry eigenvalues of discrete Laplacians \cite{WQFT2}. These are denoted by 
colored  propagator lines. The exciton propagator, written in Matsubara formalism,
is given by
\bea
\hspace*{-0.2cm}D(\bm x - \bm y, \tau_x - \tau_y) & = & 
\Lan \mbox{T} \left\{ \psi(\bm x,\tau_x) \psi^\dagger (\bm y,\tau_y) 
 \right\} \Ran_0  \label{Prop} \\
& = &  \frac 1\beta \sum_{n = -\infty}^\infty 
\int_{\bm q} \frac{\e^{\i \bm q \cdot (\bm x - \bm y) - \i
 \omega_n (\tau_x-\tau_y)}}{E(\bm q) - \i \omega_n}, \nonumber
\eea
where
\bea
\int_{\bm q} \equiv  \int_{\text{IBZ}} \frac{\d^D \bm q}{(2 \pi)^D}. \label{Omega0}
\eea
Within one-loop approximation, there are four types of diagrams. 
They can be easily evaluated
\begin{widetext}
\bea
\begin{array}{l} \vspace{0.95cm}
 {\includegraphics[scale=0.8]{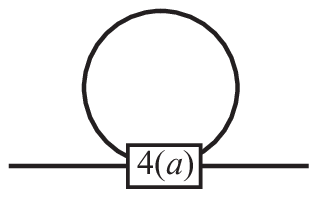}} \end{array} & = &
\begin{array}{l} \vspace{0.95cm}
 {\includegraphics[scale=0.8]{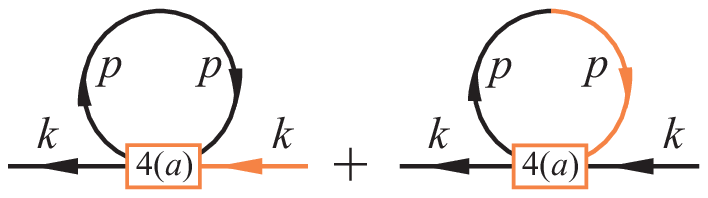}} \end{array}
=
\frac{1}{S} \frac{v_0}{2 m_0} 
\int_{\bm p} \langle n_{\bm q} \rangle_0 \; 
\left[\widehat{\bm k}^2 + \widehat{\bm p}^2 \right], \\
\begin{array}{l} \vspace{0.95cm}
 {\includegraphics[scale=0.8]{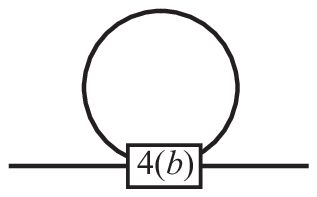}} \end{array} & = & 
\begin{array}{l} \vspace{0.95cm}
 {\includegraphics[scale=0.8]{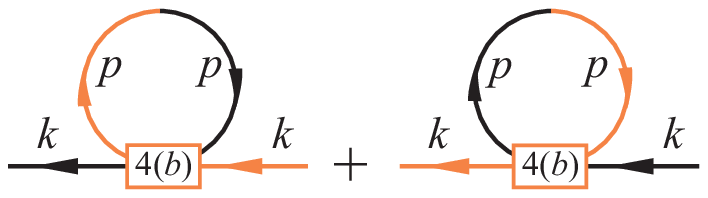}} \end{array} 
  =  -
\frac{1}{S} \frac{v_0}{2 m_0} 
 \int_{\bm p} \langle n_{\bm p} \rangle_0 \; \widehat{\bm k - \bm p}^2 ,
\eea 
\end{widetext}
with $\widehat{\bm k}^2$ defined in (\ref{KSqDef}) and $\langle n_{\bm p} \rangle_0$
denoting the free exciton Bose distribution.
Explicit expression for exciton self-energy is found by
using relation
\bea
\int_{\bm q} \Lan n_{\bm q} \Ran_0 \; \widehat{\bm p - \bm q}\; ^{2}_{(j)} = 
\int_{\bm q} \Lan n_{\bm q} \Ran_0 \left[\widehat{\bm p}_{(j)}^{\;2} + \widehat{\bm q}_{(j)}^{\;2}  
- \frac{|\bm \lambda_j|^2}{2 D} \widehat{\bm p}_{(j)}^{\;2} \; \widehat{\bm q}_{(j)}^{\;2} \right]
\nonumber
\eea
and is given by
\bea
\Sigma(\bm k) = \frac{\widehat{\bm k}^2_{(3)}}{2 m} A_3(T)
 + \frac{\widehat{\bm k}^2_{(1)}}{2 m} A_1(T) 
+\frac{\widehat{\bm k}^2_{(2)}}{2 m} A_2(T).
\eea
The temperature dependent factors $A_j(T)$ are
\bea
A_1(T) & = &\frac{a^2}{2D} \frac{|\bm \lambda_1|^2}{|\bm \lambda_3|^2} \frac{I_1^x}{I^x_3} 
v_0\int_{\bm q} \Lan n_{\bm q} \Ran_0 \widehat{\bm q}_{(1)}^{\;2}, \nonumber\\
A_2(T)& = &\frac{b^2}{2D} \frac{|\bm \lambda_2|^2}{|\bm \lambda_3|^2} \frac{I_2^x}{I^x_3} 
v_0\int_{\bm q} \Lan n_{\bm q} \Ran_0 \widehat{\bm q}_{(2)}^{\;2}, \nonumber \\
A_3(T)& = &\frac{2|\bm \lambda_3|^2}{2D} 
v_0\int_{\bm q} \Lan n_{\bm q} \Ran_0 \widehat{\bm q}_{(3)}^{\;2}. \label{AjT}
\eea
They are dimensionless quantities that  capture the effects of 
exciton-exciton interactions in two-level
system by renormalizing exchange integrals $I_j^x \to I_j^x(T)$.  
Finally, renormalized exciton energies read \cite{Wen}
\be
E_{\text{R}} (\bm k) = E(\bm k) - \Sigma (\bm k),
\ee
and the influence of exciton-exciton interactions at one-loop order can
be seen from Fig. \ref{ATemp}.
\bc
\ec


\bc
\begin{figure} 
\includegraphics[width=\columnwidth]{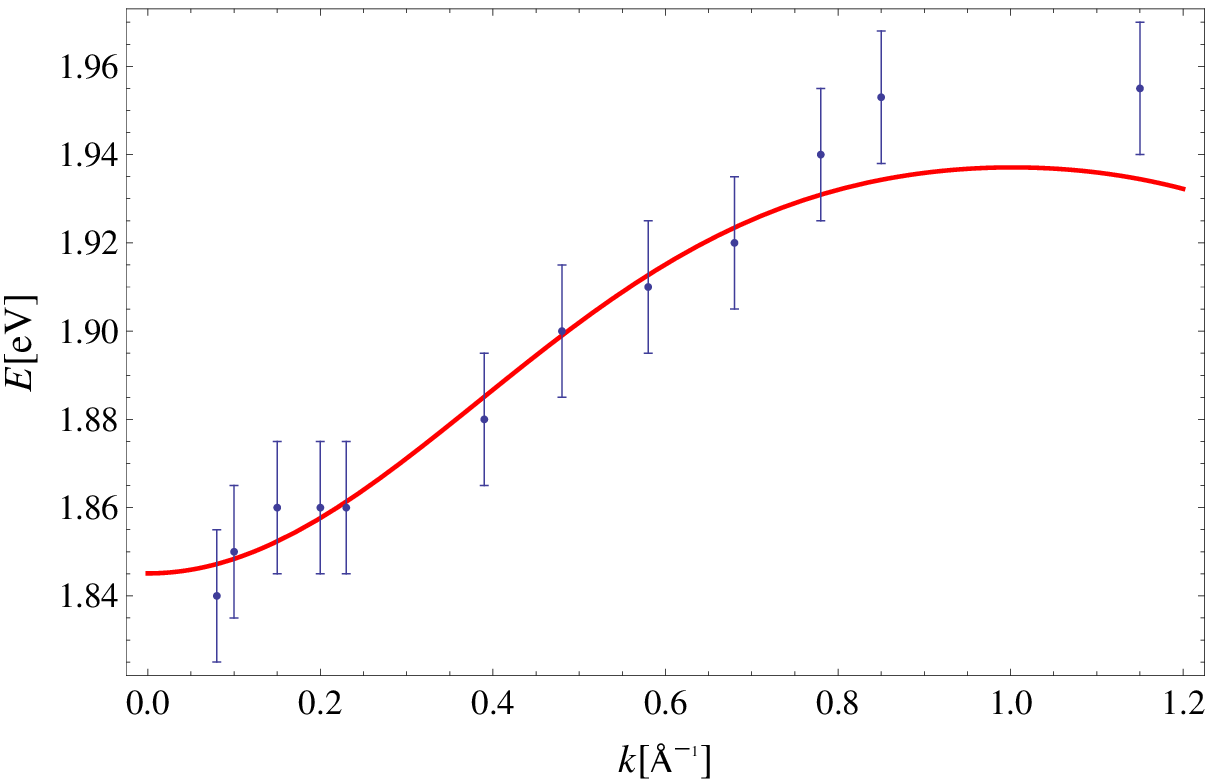} 
{\caption{\label{Fig100JPC}
Exciton dispersion along $(100)$ direction. Experimental data are taken 
from \cite{20K}. Theoretical curve is obtained for: $\Delta = 1.90 \text{eV}$ \cite{183},
$I^x_1=5.7 \text{meV}$, $I^x_2=3.4 \text{meV}$ and $I^x_3=23.4 \text{meV}$.}}
\end{figure}
\ec

\bc
\begin{figure} 
\includegraphics[width=\columnwidth]{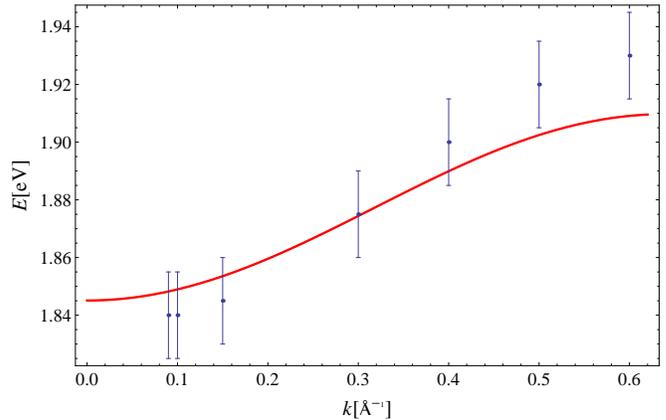}
{\caption{\label{Fig110JPC}
Exciton dispersion along $(110)$ direction. Experimental data are taken 
from \cite{20K}. Theoretical curve is obtained for: $\Delta = 1.90 \text{eV}$ \cite{183},
$I^x_1=5.7 \text{meV}$, $I^x_2=3.4 \text{meV}$ and $I^x_3=23.4 \text{meV}$.}}
\end{figure}
\ec

As noted in  Introduction, the experimental data on
exciton dispersion along $(100)$ and $(110)$ directions
 at lower temperatures are available \cite{20K}.
One can see from Figs \ref{Fig100JPC} and \ref{Fig110JPC} that
the effective model (\ref{FreeHam}), with parameters  
$\Delta = 1.90 \text{eV}$ \cite{183},
$I^x_1=5.7 \text{meV}$, $I^x_2=3.4 \text{meV}$ and $I^x_3=23.4 \text{meV}$
gives exciton dispersion in satisfying agreement with the
experimental one. The
only difference between two sets of parameters, describing
experimental data obtained at $300 \text{K}$ and $20 \text{K}$,
is the value of the parameter $\Delta$, which originates from the change of magnetic field $\mathcal{H}$ in (\ref{spinham}).

\section{Discussion} 

As seen from Fig. \ref{ATemp}, the influence of exciton-exciton
interactions is negligible at room temperatures. There are
two main reasons for this. First, the exciton gap is huge -- nearly
two orders of magnitude larger than the greatest exchange integral.
Second, excitons are derivatively coupled via interactions
that are of the type occuring in the nonlinear $\sigma$ models.
Since these interactions include Laplacians, they tend to vanish at low
energies. In fact, recent studies \cite{PRD, PTEP} have shown that
scattering amplitudes in a system governed by such interactions
disappear  as momenta of particles tend to zero. This interpretation
is similar to the one given by Dyson in his analysis of 
ferromagnetic systems \cite{Dyson1, Dyson2}. On the other hand,
as scattering amplitudes tend to zero regardless of exciton gap
(i.e. fictitious external magnetic field of corresponding
ferromagnetic system),
it contradicts to the "hard sphere" picture of exciton
dynamics from \cite{Tole}.

It is important to compare results from the present paper to 
the ones obtained by solving Bethe-Salpeter (BS) equations
for many-body electron-hole system \cite{GatuzoPRB2,GatuzoJPC}.
First, the exciton dispersion obtained here is
closer to  experimental values. This is clearly seen by comparing Figures \ref{Fig100}-\ref{Fig110}
from present paper to the results of Cudazzo et al. (see Figure 3 in \cite{GatuzoPRB2}).
Second, the exciton dispersion obtained with the help of the effective
model in the present paper is much more robust against perturbative corrections. This
may also be seen from Figs 3 and 4 from  \cite{GatuzoPRB2}: The  
dispersion  obtained using flat HOMO-LUMO bands yields
exciton dispersion close to $3.5 \text{eV}$, while
those obtained 
using HOMO-LUMO with full dispersion are between $1.77 \text{eV}$ and $1.8 \text{eV}$ along (100) direction.

To conclude, we have analyzed  the exciton dispersion
in  pentacene relying on the correspondence between Pauli (\ref{paulham}) and
Heisenberg (\ref{spinham}) Hamiltonians. By fitting exchange integrals
to the experimental data, we have obtained exciton dispersion
that possesses the same periodicity as experimentally observed
one. Also, our results provide an indirect confirmation that
2D model is indeed a minimal one that describes available experimental
data on exciton dispersion.
Further, we have shown that exciton-exciton interactions produce
negligible effects to the one loop order.
Because of that, we suggest that the influence of CT excitons
in pentacene, which needs to be taken into account, 
may be less important than indicated in previous studies. 
It would be interesting
to see experimental data on exciton dispersion along $\bm c^*$ axis and how 
this data would fit into existing models.  Also, it is important to understand how
to improve calculations based on BS equations to reach better agreement 
with experimental data on exciton dispersion and to test that approach for
all four directions within Brillouin zone considered in \cite{breakdown}.
 Therefore, further experimental and theoretical work is 
necessary before drawing final conclusion regarding the influence of
CT excitons in pentacene.

\section*{Acknowledgments}
We are grateful to R. Schuster for sharing experimental 
data from \cite{breakdown} and for helpful discussions. 
We also wish to thank the referees for their helpful comments.
This work was supported by the Serbian Ministry of
Education and Science under Contract No. OI-171009
and by the Provincial Secretariat for High Education and Scientific 
Research of Vojvodina (Project No. APV 114-451-2201).

\bibliography{Refs}

\end{document}